\documentclass[usletter,11pt]{article}
\usepackage{jheppub}

\usepackage{listings}
\usepackage{tikz}
\usepackage{orcidlink}
\usepackage{slashed}  
\usepackage{subcaption}
\usepackage{colonequals}
\usepackage{rotating}
\usepackage{datetime2} 
\usepackage{multirow}
\usepackage{xspace}
\usepackage{datetime}
\usepackage[framemethod=tikz]{mdframed}
\usepackage{pgf,tikz}
\usepackage{mathrsfs}
\usetikzlibrary{arrows}
\usepackage{pgf,tikz}
\usetikzlibrary{arrows.meta}
\usepackage{graphicx,tipa}

\usepackage{hyperref}
\usepackage[braket, qm]{qcircuit}
\usepackage{lmodern}
\usepackage[x11names]{xcolor}
\usepackage{framed}
\colorlet{shadecolor}{LavenderBlush2} 
\colorlet{framecolor}{Red1}
\usepackage[framemethod=tikz]{mdframed}
\usepackage{soul}

\definecolor{dkgreen}{rgb}{0,0.6,0}
\definecolor{gray}{rgb}{0.5,0.5,0.5}
\definecolor{mauve}{rgb}{0.58,0,0.82}
\title{Quantum simulation of massive Thirring and Gross--Neveu models for arbitrary number of flavors}

\author[a]{Bojko N.~Bakalov~\orcidlink{0000-0003-4630-6120},}
\author[b]{João C.~Getelina~\orcidlink{0000-0002-1924-9813},}
\author[b]{Raghav G.~Jha~\orcidlink{0000-0003-2933-0102},}
\author[b]{Alexander F.~Kemper~\orcidlink{0000-0002-5426-5181},} 
\author[cdb]{Yuan Liu~\orcidlink{0000-0003-1468-942X}}

\affiliation[a]{Department of Mathematics, North Carolina State University, Raleigh, North Carolina 27695, USA}
\affiliation[b]{Department of Physics and Astronomy, North Carolina State University, Raleigh, North Carolina 27695, USA}
\affiliation[c]{Department of Electrical and Computer Engineering,
North Carolina State University, Raleigh, North Carolina 27695, USA}
\affiliation[d]{Department of Computer Science,
North Carolina State University, Raleigh, North Carolina 27695, USA}

\emailAdd{bnbakalo@ncsu.edu}
\emailAdd{jdeandr2@ncsu.edu}
\emailAdd{raghav.govind.jha@gmail.com}
\emailAdd{akemper@ncsu.edu}
\emailAdd{q\_yuanliu@ncsu.edu}

\abstract{
The study of fermionic quantum field theories is an important problem for realizing the standard model of particle physics on a quantum computer. As a step towards this goal, 
we consider the massive Thirring and Gross--Neveu models with arbitrary number of fermion flavors, $N_f$, discretized on a spatial one-dimensional lattice of size $L$ in the Hamiltonian formulation. We compute the gate complexity using the higher-order product formula and using block-encoding/qubitization and quantum singular value transformations in the limit of large $N_f$ and $L$. We also prepare the ground states of both models with excellent fidelity for system sizes up to 20 qubits with $N_f = 1,2,3,4$ using the adaptive-variational quantum imaginary time algorithm. In addition, we also classify the dynamical Lie algebras of these relativistic fermionic models and show that they belong to the same isomorphism class. Our work is a concrete step towards the quantum simulation of real-time dynamics of large $N_f$ fermionic quantum field theories models relevant for chiral symmetry breaking, understanding dimensional transmutation, and exploring the conformal window of field theories on near-term and early fault-tolerant quantum computers.
}

\begin{document}
\maketitle
\flushbottom

\section{Introduction}
\label{sec:sec_intro}

The theory of strong interactions, known as quantum chromodynamics (QCD), has achieved great success in explaining the structure of matter in the last six decades. A proper understanding of QCD in the non-perturbative regime is essential for wide ranges of phenomena that occur in nature, such as nuclear structure, confinement of quarks, and the composition and microscopic behavior of neutron stars. In particular, many of the most pressing questions involve real-time dynamics and matter at finite baryon density.
Lattice methods are powerful tools for studying non-perturbative physics, but are restricted to Euclidean (imaginary) time and real lattice actions. In lower dimensions, four-fermion interaction models have been studied using classical methods, including tensor networks~\cite{Pannullo:2019bfn, Banuls:2019hzc, Roose:2020znu,Asaduzzaman:2022bpi}, but this approach is not efficient in higher dimensions. 

One reason why the real-time dynamics of a lattice Hamiltonian with a large number of flavors $N_f$ can be important is for understanding chiral symmetry breaking (CSB)~\cite{PhysRevResearch.2.023342, Wellegehausen:2017goy}, which is the 
mechanism that explains how quarks (e.g., protons and neutrons) get almost all of their mass. Depending on whether one is in the chirally broken or symmetric phase, the time-evolved vector current after a global quench for different numbers of fermion flavors will behave differently, allowing, in principle, the critical flavor number to be computed\footnote{The critical flavor number $N_{\text{cr}}$ is defined such that chiral symmetry breaking occurs only for $N_f \le N_{\text{cr}}$.}. Quantum computing is an alternative computational approach to understanding finite-density and real-time dynamics of QCD. Motivated by this possibility, several groups have previously studied models relevant to fermionic quantum field theories within the framework of quantum computation~\cite{Jordan:2014tma, HamedMoosavian:2017koz, Mishra:2019xbh, Asaduzzaman:2022bpi, Gong:2024iqu}. 

Most of the existing work related to simulating QCD on a quantum computer has focused on lower dimensions with a small or fixed number of fermion flavors~\cite{Zohar:2012xf, Wiese:2014rla, Kan:2021xfc}. However, in QCD, there are six types (or flavors, denoted $N_f$) of quarks. All these flavors have been detected in nature and are theoretically established within the framework of the Standard Model of particle physics. Any complete future quantum computational study of QCD must therefore account for multiple quark flavors. Studying finite-$N_f$ and large-$N_f$ limits of fermionic quantum field theories is thus crucial for understanding various real-time/non-perturbative features of QCD and its limitations. 
In this paper, we provide a detailed quantum computational study of the large $N_f$ limit of fermionic quantum field theories (QFTs) in 1+1-dimensions. 

Although several works have considered QCD in 1+1-dimensions~\cite{Than:2024zaj, Yang:2025edn}, our paper focuses on the strict large $N_f$ limit for the Hamiltonian simulation resource estimate and on $N_f=4$ for the ground state preparation of certain fermionic models relevant for the matter sector of QCD. 
For a wide range of models relevant to the theory of strong interactions and other quantum field theories, the main hurdle is the lack of a suitable method for studying real-time dynamics on a discretized space-time lattice. One approach is to go beyond the paradigm of classical computing and instead use quantum computing to attack this problem. Quantum computers will likely offer the possibility to study real-time dynamics of quantum field theories in higher dimensions, and thus improve our understanding of fundamental interactions, including nuclear processes in QCD. However, this problem is still beyond the reach of current capabilities and requires several more orders of magnitude of resources than currently available~\cite{Kan:2021xfc, Tong:2021rfv}.

Contrary to the Lagrangian approach to QCD, the Hamiltonian approach requires more effort, because one has to impose the gauge constraints and locality explicitly by building the physical Hilbert space and operators that act on it, while this is naturally embedded in the path integral formulation.
This has led to various formulations and ways of constructing a suitable Hamiltonian lattice gauge theory that captures the physics of strong interactions~\cite{Kogut:1974ag, Kemper:2025ldr}. As a step towards this long-term goal, we begin by considering two well-studied toy models of relativistic fermions in 1+1 dimensions with a large number of flavors $N_f$. We perform quantum simulation of the ground-state preparation and calculate the resources required for the Hamiltonian simulation using both higher-order product formulas and block-encoding/quantum singular value transformation (QSVT).

The toy models we consider in 1+1-dimensions exhibit a number of features similar to 3+1-dimensional QCD, including negative $\beta$ function (running of coupling with energy scale) leading to asymptotic freedom, and other interesting phenomena such as chiral symmetry breaking and dynamical mass gap generation. The simplest class of interacting fermionic quantum field theories (QFTs) is based on four-fermion interactions in 1+1 dimensions. The first such model was introduced by Thirring~\cite{Thirring1958} and was extended to another four-fermion model by Gross and Neveu~\cite{Gross1974}. The difference between the Thirring model and the Gross--Neveu (GN) model is that, while the Thirring model involves a vector-current interaction term $ (\overline{\psi} \gamma^\mu \psi)^2 $, the GN model has a scalar--scalar interaction of the form $ (\overline{\psi} \psi)^2 $, where $ \psi $ represents the fermion field (Dirac spinor), which is a two-component object in 1+1 dimensions. For the special case of $N_f=1$, both models are equivalent up to a constant factor due to the Fierz identity\footnote{This also results in different sign for the interaction term in the Thirring and Gross--Neveu model.} given by $-2(\bar{\psi} \psi)^{2} = (\overline{\psi} \gamma^{\mu} \psi)^{2}$ in the absence of a pseudoscalar bilinear term~\cite{Martins:2024dag}. The massive Thirring model with $N_f=1$ is a special case, since it can be mapped to a bosonic model, i.e., the sine--Gordon model~\cite{Coleman1975, Mandelstam1975}. 
This well-known duality between the single-flavor massive Thirring model and the sine--Gordon model, involving solitonic degrees of freedom in $1+1$ dimensions, is an example of ``bosonization''. Due to this, Thirring model is exactly solvable and has been extensively studied~\cite{Coleman1975, Korepin:1979qq}. However, in general, the GN model is not solvable for finite $N_f$.

We also present results for the classification of dynamical Lie algebras (DLA) for fermionic Hamiltonians that correspond to relativistic QFTs in the continuum limit. Recall that the DLA is defined as the Lie algebra obtained by taking all real linear combinations and nested commutators of the terms of the Hamiltonian~\cite{DAlessandro2021}. The significance of the DLA is that the time evolution of the system is given by elements of the associated Lie group. The DLA determines the set of reachable states of the system and its controllability~\cite{DAlessandro2021}, as well as the trainability of parameterized quantum circuits~\cite{Ragone:2023qbn}, and is also useful for fast-forwarding of certain Hamiltonians~\cite{Kokcu:2021ctj, Alsheikh:2025gbh}. We sketch out some details about DLAs in Appendix~\ref{sec:dla_appendix} and refer to Refs.~\cite{DAlessandro2021, Schirmer:2002kuu, Khaneja:2000stb,Goh:2023kcm,Wiersema:2023txu, Kazi:2025qpb} for further reading. DLAs associated to spin systems were classified in \cite{Wiersema:2023txu}, and more generally, those generated by Pauli strings were classified in \cite{Aguilar:2024dnz}. We identify the DLA generated by the Pauli terms of fermionic Hamiltonians with one of the classes in \cite{Aguilar:2024dnz} in  Table~\ref{tab:DLA_all} of Sec.~\ref{sec:DLA_class}. 

\begin{figure}
    \centering
    \includegraphics[width=0.8\linewidth]{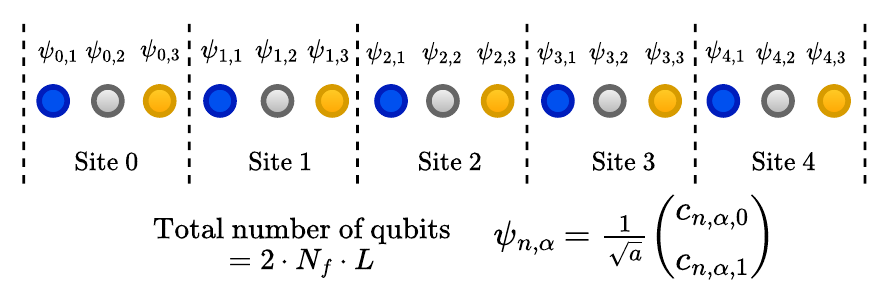}
    \caption{Five-site lattice ($L=5$) with three flavors ($N_f =3$) of fermions (different colors) at each physical lattice site. Each blob (i.e., Dirac spinor) is represented by two qubits. The lattice spacing is denoted as $a$.} 
    \label{fig:cartoon1}
\end{figure} 
The paper is organized as follows. In Sec.~\ref{sec:qubit_form}, we formulate the Thirring and Gross--Neveu models for arbitrary number of fermionic flavors, $N_f$, in a form suited for quantum computation. Then, we present our results in Table~\ref{tab:table_AVQITE} for the ground state preparation using AVQITE in Sec.~\ref{sec:AVQITE_GS_prep} for both models for system size up to 20 qubits and $N_f=4$ with excellent fidelity. In addition, using the AVQITE prepared ground state, we compute the static equal-time fermion bilinear. In Sec.~\ref{sec:resource_costs}, we present the gate complexity of simulating the four-fermion models using higher-order product formulas and using a block-encoding constructed via linear combination of unitaries (LCU) followed by quantum singular value transformations (QSVT) for large number of lattice sites, $L$ and flavors, $N_f$. In Sec.~\ref{sec:DLA_class}, we discuss the classification of dynamical Lie algebra (DLA) of these models, which are essential to understanding the landscape of variational quantum optimization and barren plateaus. We conclude the paper with a summary and discussion in Sec.~\ref{sec:summary_fd}. We provide additional details about Hamiltonian construction, AVQITE procedure, dynamical Lie algebras, and the gate counts in Appendix~\ref{sec:cx+t_appendix} for the interested reader. 

\section{\label{sec:qubit_form}Four-fermion models and qubit formulation}
The Lagrangian for single flavor massive Thirring model is given by:
\begin{align}
    \mathcal{L}_{\mathrm{Thirring}} = i\overline{\psi}( \slashed{\partial} - m) \psi - g (\overline{\psi} \gamma^{\mu}\psi)(\overline{\psi} \gamma_{\mu}\psi),
    \label{eq:Lagrangian_Thirring}
\end{align}
and for single flavor massive Gross--Neveu by: 
\begin{align}
    \mathcal{L}_\mathrm{GN} = i\overline{\psi}( \slashed{\partial} - m) \psi + g (\overline{\psi}\psi)^2,
    \label{eq:Lagrangian_GN}
\end{align}
where we denote $\slashed{\partial}=\gamma^{\mu}\partial_{\mu}$, according to the standard Feynman's slash notation, and $\overline{\psi} = \psi^{\dagger}\gamma^{0}$ is the Dirac adjoint of two-component spinor $\psi = \begin{pmatrix} \psi_0 \\ \psi_1 \end{pmatrix} 
$. For the gamma matrices\footnote{We use the mostly minus/West-coast metric i.e., metric signature $(+ -)$ where $\gamma^{0}$ is Hermitian and $\gamma^1$ is anti-Hermitian.}, we choose the standard representation (Dirac representation) where $\gamma^{0} = \sigma_z =  Z = \begin{pmatrix} 1 & 0 \\ 0 & -1 \end{pmatrix}$ and $\gamma^{1} = i \sigma_y = i Y = \begin{pmatrix} 0 & 1 \\ -1 & 0 \end{pmatrix}$.
In this paper, our focus is to initiate a quantum computing study of the massive Thirring/GN model with $N_f$ flavors, where the interaction term is  $(\overline{\psi}_{\alpha} \gamma^{\mu} \psi_{\alpha})^{2}$ for the Thirring model, and $(\overline{\psi}_{\alpha} \psi_{\alpha})^{2}$ for the Gross--Neveu model. The sum over flavors $\alpha$ is always implied if not explicitly written, i.e., $\overline{\psi}_{\alpha} \gamma^{\mu} \psi_{\alpha} = \sum_{\alpha=1}^{N_f} \overline{\psi}_{\alpha} \gamma^{\mu} \psi_{
\alpha}$. We choose the same mass for all flavors, i.e., $m_\alpha = m$. 

Starting with the generalization 
of the Lagrangian given by Eq.~\eqref{eq:Lagrangian_Thirring} to the multi-flavor case, we can write down the Hamiltonian of the massive Thirring model for $N_f$ flavors with open boundary conditions as: 
\begin{align}
H_\mathrm{Thirring} =& -i\sum_{n=0}^{L-2} \sum_{\alpha = 1}^{N_f}\left(\overline{\psi}_{n,\alpha} \gamma^{1}\partial_{1}\psi_{n+1,\alpha} \right) 
+ m \sum_{n=0}^{L-1} \sum_{\alpha = 1}^{N_f} \overline{\psi}_{n, \alpha} \psi_{n,\alpha} \nonumber\\
&+ g \sum_{n=0}^{L-1} \sum_{\alpha=1}^{N_f}  (\overline{\psi}_{n,\alpha}\gamma^{\mu}\psi_{n,\alpha}) \cdot (\overline{\psi}_{n, \alpha}\gamma_{\mu}\psi_{n, \alpha}).
\end{align} 
In order to consider the lattice Hamiltonian on discretized spatial lattice, we rewrite the two-component fermion field $\psi(x)$ at site $x$ as:
\begin{equation}
 \psi(x) = \begin{pmatrix} \psi_0 \\ \psi_1 \end{pmatrix} = 
\begin{pmatrix} c_{0} \\ c_{1} \end{pmatrix},   
\end{equation}
with the lattice spacing being set to $a=1$. For the kinetic term, we need the derivative on the spatial lattice $\partial_1$ for which we use the standard (symmetric) finite difference~\cite{Susskind:1976jm}:
\begin{equation}
 \partial_{1} \psi_n = \frac{1}{2a} (\psi_{n+1} - \psi_{n-1}).    
\end{equation}
This choice of naive lattice derivative is known to lead to the `fermion doubling' problem consisting of $2^d$ spurious zero modes where $d$ is number of spatial dimensions. Most of the lattice studies either use the Wilson approach~\cite{Wilson:1974sk} or the staggered fermion approach~\cite{Susskind:1976jm}. In the former approach, a term is added to the Hamiltonian which lifts the doublers while in the staggered approach, one adopts a sub-lattice structure keeping the different components of Dirac spinor on even/odd sites. In 1+1-dimensions, staggered approach leads to complete removal of the doublers. In practical simulations of this model in the future, it will be essential to add a Wilson-like term~\cite{Wilson:1974sk, Jordan:2014tma}. However, for this work we do not include this additional term.

Given a four-fermion Hamiltonian defined on $L$ sites, one requires $2N_f$ qubits per site, giving a total of $n = 2N_fL$ qubits. The Hamiltonian is a sum of three terms: the kinetic term $H_k$, the mass term $H_m$, and the four-Fermi term, $H_{\text{FF}}$. Only the four-Fermi term is different between the two models. The kinetic term is:
\begin{align}
    H_{k} &= -i \sum_{n=0}^{L-2} \sum_{f=0}^{N_f-1}  
    \overline{\psi}_{n,f} \,\gamma^{1} \partial_{1} \,\psi_{n,f}, \nonumber \\
    &= -\frac{i}{2} \sum_{n=0}^{L-2} \sum_{f=0}^{N_f-1}  
    \Big[\begin{pmatrix} c_{n,f,0}^{\dagger} & c_{n,f,1}^{\dagger} \end{pmatrix} \underbrace{\begin{pmatrix} 0 & 1 \\
   1 & 0 \end{pmatrix}}_{\gamma^0 \gamma^1 = X} \begin{pmatrix} c_{n+1,f,0} \\ c_{n+1,f,1} \end{pmatrix} \,- \begin{pmatrix} c_{n+1,f,0}^{\dagger} & c_{n+1,f,1}^{\dagger} \end{pmatrix}\,X\, \begin{pmatrix} c_{n,f,0} \\ c_{n,f,1} \end{pmatrix}\Big], \nonumber \\ 
    &=- \frac{i}{2} \sum_{n=0}^{L-2} \sum_{f=0}^{N_f-1} \Big[
    c_{n,f,0}^\dagger c_{n+1,f,1}
    + c_{n,f,1}^\dagger c_{n+1,f,0}
    - c_{n+1,f,0}^\dagger c_{n,f,1}
    - c_{n+1,f,1}^\dagger c_{n,f,0}
\Big].
\label{eq:kinetic_term_H}
\end{align}
We now consider the mass term which, like the kinetic term, is the same for both models and given by: 
\begin{equation}
 H_{m} = m  \sum_{n=0}^{L-1} \sum_{f=0}^{N_f-1} 
\left[ 
    \left( c_{n,f,0}^\dagger c_{n,f,0} 
    - c_{n,f,1}^\dagger c_{n,f,1} \right) 
\right]. 
\label{eq:mass_term_H}
\end{equation}
The two models differ in the interaction term. For the Thirring model, we have: 
\begin{align}
   H_{\text{FF, Thirring}} &= g \sum_{n=0}^{L-1}  \Bigg[\sum_{\alpha = 1}^{N_f} \psi^{\dagger}_{n, \alpha} \gamma^0 \gamma^\mu \psi_{n,\alpha} \Bigg] \cdot \Bigg[\sum_{\alpha = 1}^{N_f} \psi^{\dagger}_{n, \alpha} \gamma^0 \gamma_\mu \psi_{n,\alpha} \Bigg],  \nonumber \\ 
   &= g \sum_{n=0}^{L-1} \Bigg( \Bigg[\sum_{\alpha = 1}^{N_f} \psi^{\dagger}_{n,\alpha}\,\mathbb{I}\,\psi_{n,\alpha} \Bigg]^{2} - \Bigg[\sum_{\alpha = 1}^{N_f} \psi^{\dagger}_{n,\alpha}\,X\,\psi_{n,\alpha} \Bigg]^{2}\Bigg), \nonumber \\ 
   &= g \sum_{n=0}^{L-1} \Bigg(\Bigg[\sum_{\alpha = 1}^{N_f} \Big(c^\dagger_{n, \alpha, 0} c_{n, \alpha, 0}
+ c^\dagger_{n, \alpha, 1} c_{n, \alpha, 1}\Big)\Bigg]^{2} -  \Bigg[\sum_{\alpha = 1}^{N_f} \Big(c^\dagger_{n, \alpha, 0} c_{n, \alpha, 1}
 c^\dagger_{n, \alpha, 1} c_{n, \alpha, 0}\Big)\Bigg]^{2}\Bigg),
\label{eq:int_Thirring}
\end{align}
while for the Gross--Neveu (GN) model it is:
\begin{align}
H_{\text{FF, Gross--Neveu}} &= -g \sum_{n=0}^{L-1} \Bigg[\sum_{\alpha = 1}^{N_f} \Big(\psi^{\dagger}_{n,\alpha} \gamma^{0} \psi_{n,\alpha} \Big)\Bigg]^{2}, \nonumber \\ 
&= -g \sum_{n=0}^{L-1} \Bigg[\sum_{\alpha=1}^{N_f} \Big(c^\dagger_{n, \alpha, 0} c_{n, \alpha, 0} - c^\dagger_{n, \alpha, 1} c_{n, \alpha, 1}\Big)\Bigg]^{2}.  
\label{eq:int_GN} 
\end{align}
The total Hamiltonian is the sum of Eqs.~\eqref{eq:kinetic_term_H}, \eqref{eq:mass_term_H}, \eqref{eq:int_Thirring} for the Thirring model, and Eqs.~\eqref{eq:kinetic_term_H}, \eqref{eq:mass_term_H}, \eqref{eq:int_GN} for the Gross--Neveu model. In order to make this problem amenable to quantum computing, we encode the Dirac spinors using the Jordan--Wigner (JW) transformation. Using this fermion-qubit mapping, we can rewrite the full Hamiltonian completely in terms of Pauli matrices for any number of flavors utilizing a total of $2N_{f}L$ qubits for a spatial lattice of $L$ sites. As a simple explicit example, we provide the Hamiltonian for $N_f = L = 2$ with open boundary condition in Appendix~\ref{sec:sampleH_appendix}.

\section{\label{sec:AVQITE_GS_prep}Ground state preparation using AVQITE and static fermion correlator}

In this section, we present our results on applying a quantum algorithm to prepare the ground state of the four-fermion models discussed in the previous section. We also show results for a fermion-condensate two-point correlator using the prepared ground state and compare to exact results.  

\subsection{AVQITE preparation}
The state-preparation method that we have chosen is the adaptive-variational quantum imaginary-time evolution (AVQITE)~\cite{Gomes:2021ckn,Getelina:2023nwx,Getelina:2023yrf}. AVQITE builds on the quantum imaginary-time evolution~\cite{Motta:2020qite} and its variational counterpart~\cite{McArdle:2019vqite}, by allowing the underlying parameterized circuit to expand iteratively as needed, thus saving quantum resources and producing shallower circuits than other imaginary-time approaches, making it more suitable for near-term quantum simulations. In addition, after using AVQITE to prepare the ground state, one can employ the same adaptive-variational principles to perform the real-time dynamics~\cite{Yao:2021avqds}. Additional details about AVQITE can be found in Appendix~\ref{sec:qite_avqite_appendix}.

The cornerstone of AVQITE is the so-called McLachlan's variational principle~\cite{McLachlan:1964,Yuan:2018jdl}, which consists of finding an optimal parameterized unitary $U_{\tau}(\boldsymbol{\theta})$ that closely resembles the path taken by the imaginary-time evolution of the system, the latter being governed by the non-unitary operator $\exp(-\tau H)$, with $\tau$ being the imaginary time and $H$ the system Hamiltonian. The adaptive part of this procedure stems from iteratively appending operators to the parameterized unitary $U_{\tau}(\boldsymbol{\theta})$, until the minimization criterion is satisfied. These unitary operators are drawn from a predefined operator pool, which is in general composed of Pauli strings
with maximum weight $k$.

\begin{table}[h!]
\centering
  \setlength{\tabcolsep}{4pt}     
  \renewcommand{\arraystretch}{1.4}%
\begin{tabular}{c|c| c| c | c| c}
\hline
\textbf{($L, N_f, g$)} & \textbf{Model} & \textbf{Operator pool} & \textbf{$E_{\text{Exact}}$} & \textbf{$\Delta$ (in \%)} & $\mathbf{F}$ \\
\hline
(8,\,1,\,0.2) & GN &  \{YY, YYZ\} & -7.640 & $0.04$ & 0.99 \\
\hline 
(9,\,1,\,0.2) & GN & \{YY, YYZ\} & -8.626 & $0.03$ & 0.99 \\
\hline 
(10,\,1,\,0.2) & GN & \{XY, YYZ\} & -9.612 &  $0.03$ & 0.99 \\
\hline
(8,\,1,\,0.2) & Thirring & \{YY, YYZ\} & -5.733 &  0.05 & 0.99  \\
\hline 
$\star$ (9,\,1,\,0.2) & Thirring & \{YY, YYZ\} & -6.477 &  0.04 & 0.99  \\
\hline
(10,\,1,\,0.2) & Thirring & \{YY, YYZ\} & -7.222&  0.035 & 0.99 \\
\hline \hline 
$\star$ (4,\,2,\,0.2) & GN & \{YY, YYZ\} & -8.478 &  $0.02$ & 0.99 \\
\hline
(5,\,2,\,0.2)& GN & \{XY, YYZ\} & -10.695 &  $0.05$ & 0.99 \\
\hline 
(4,\,2,\,0.05) & Thirring & \{YY, YYZ\} & -5.570 &  0.28 & 0.99   \\
\hline
(5,\,2,\,0.05) & Thirring & \{YY, YYZ\} & -7.092 &  0.74 & 0.98   \\
\hline \hline 
(3,\,3,\,0.2) & GN & \{XY, YYZ\} & -10.875 &  $0.02$ & 0.99 \\
\hline
(3,\,3,\,0.033) & Thirring & \{XY, YYZ\} & -5.971 &  0.24 & 0.99  \\
\hline \hline 
(2,\,4,\,0.2) & GN & \{XY, YYZ\} & -10.925 &  $0.005$ & 0.99  \\
\hline
(2,\,4,\,0.025) & Thirring & \{XY, YYZ\} & -4.970 &  0.15 & 0.99   \\
\hline
\end{tabular}
\caption{\label{tab:table_AVQITE}Ground state energy error $\Delta = |E_{\text{AVQITE}} - E_{\text{Exact}}|/E_{\text{Exact}}$ (in \%)  and fidelity $F =  |\langle \psi_{\text{Exact}} | \psi_{\text{AVQITE}}\rangle|^2$ from the approximated ground state using AVQITE
and exact diagonalization for Gross–Neveu (GN) and Thirring models for various $L$, $N_f$, and $g$, with fixed $m = 0.5$ in Eqs.~\eqref{eq:Lagrangian_Thirring} and~\eqref{eq:Lagrangian_GN}. The number of qubits describing the system is $n = 2N_fL$, and the operator pool grows as $O(n^3)$. The initial state for all simulations is the Néel state, i.e., $|\psi_{\text{start}}\rangle = \ket{01}^{\otimes N_fL}$. The convergence of the ground state energy and fidelity for datasets marked by $\star$ are shown in Fig.~\ref{fig:AVQITE_results1}}
\end{table}

Constructing an efficient operator pool is a key part of AVQITE implementations, as an operator pool that is too large becomes a major bottleneck in the algorithm performance. In our simulations, we have considered all-to-all connected Pauli operators of the form $\{X_{i}Y_{j}, Y_{i}Y_{j}Z_{k}\}$ or $\{Y_{i}Y_{j}, Y_{i}Y_{j}Z_{k}\}$, with $i\ne j\ne k \in [1,2N_fL]$, which have yielded excellent fidelity and energy-error values with respect to the exact diagonalization results for all numbers of flavors and system sizes considered here. 
We tried two different operators of weight two to explore any possible computational gains and found no noticeable difference in the attained fidelity and circuit depth.
In Table~\ref{tab:table_AVQITE}, we present the fidelity and the ground state energy error estimates obtained using the AVQITE algorithm for a variety of lattice sizes $L$, fermion flavor numbers $N_f$, and coupling strengths $g$, for both the Thirring and Gross-Neveu models. We obtain two 9s of fidelity for nearly all cases, and the relative errors in the ground state energies are always below 1\%. 

\begin{figure}
    \centering
    \includegraphics[width=0.75\linewidth]{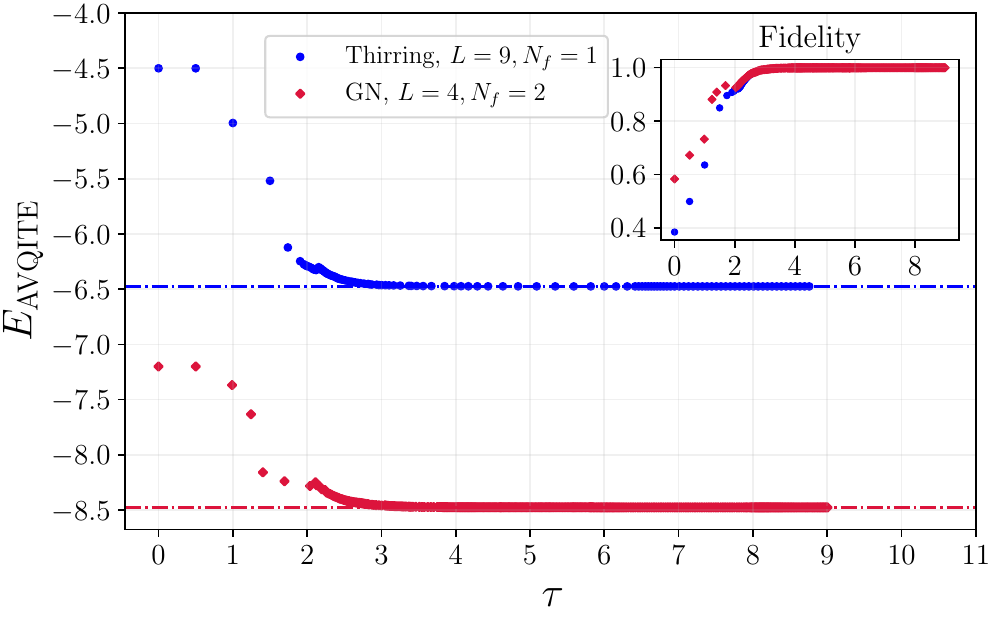}
    \caption{Convergence of the AVQITE algorithm compared to the exact ground state energy (dashed lines) and fidelity (inset) for two representative examples from the set of simulations detailed in Table~\ref{tab:table_AVQITE}.}
    \label{fig:AVQITE_results1}
\end{figure}

In Fig.~\ref{fig:AVQITE_results1}, we provide two representative examples of our simulations, one for the Thirring and the other for the GN model. In both of these cases, the ground-state energy of the approximate state quickly converges to the exact value. For the most challenging case we have considered in our simulations (determined by the infidelity, see Table~\ref{tab:table_AVQITE}), i.e., $L=5$ and $N_f=2$, which corresponds to an $n = 2N_fL=20$ qubits system, the number of two- and three-qubit operators needed in the final ansatz is 315 (454) and 325 (495), respectively, for the Thirring (GN) model. Notably, even though the number of operators in the selected pool is  $n (n - 1)/2  + n (n - 1) (n - 2)/2$, which corresponds to a scaling of $O(n^3)$, we find that for most of our simulations the final circuit consists of only $O(n^2)$ operators, as illustrated by the two- and three-local operator count being around 400 for the 20-qubit systems discussed above. 
Therefore, even though we start with $O(n^3)$ operators in the pool, we achieve excellent fidelity and small ground-state energy relative error with a significantly smaller pool size $O(n^2)$. We leave the study of a more informed selection of the operator pool for future work.

In addition to the operator pool size, the overlap of the initial state with the actual ground state can also affect the performance of AVQITE. In our simulations, we find that the overlap between the exact ground state and the initial AVQITE state decreases as we increase the system size $n$; however, the decrease is $O(\text{polylog}(1/n))$, which is better than the exponential suppression expected from a naive estimate. We show this dependence in Fig.~\ref{fig:overlap}.
For the models considered in this work, we also used ADAPT-VQE approach to ground state preparation but found that AVQITE worked consistently better. From our results in Sec.~\ref{sec:DLA_class}, it seems that variational optimization of these models will suffer from the barren plateau problem. However, with AVQITE up to 20 qubits, we did not run into this problem. We leave detailed study of how the exponential dimension of the DLA affects AVQITE performance for a future work.

\subsection{Fermion condensate}

\begin{figure}[t]
    \centering

    \begin{subfigure}[t]{0.48\linewidth}
        \centering
        \includegraphics[width=\linewidth]{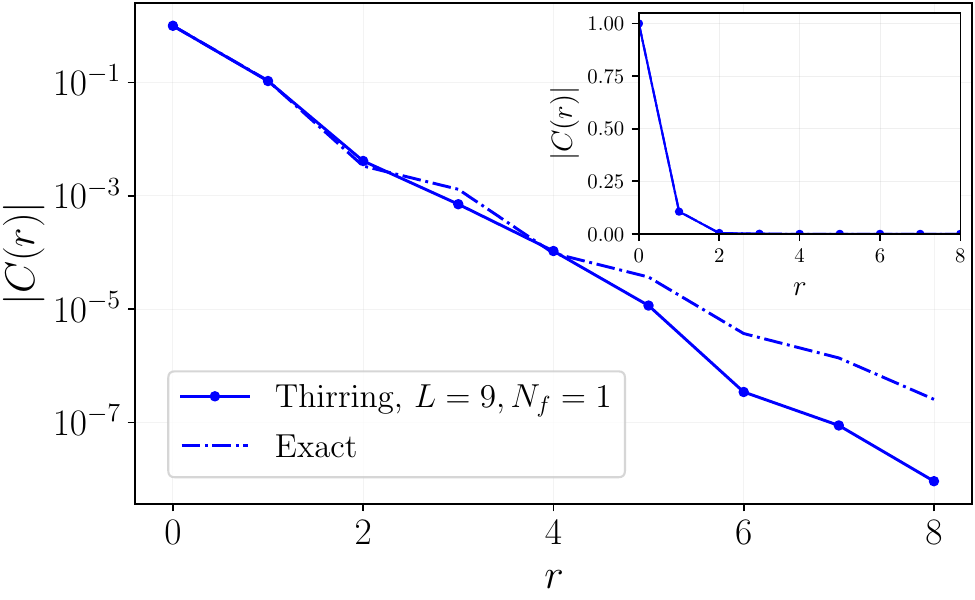}
        \caption{Thirring model}
        \label{fig:corr_AVQITE_thirring}
    \end{subfigure}\hfill
    \begin{subfigure}[t]{0.48\linewidth}
        \centering
    \includegraphics[width=\linewidth]{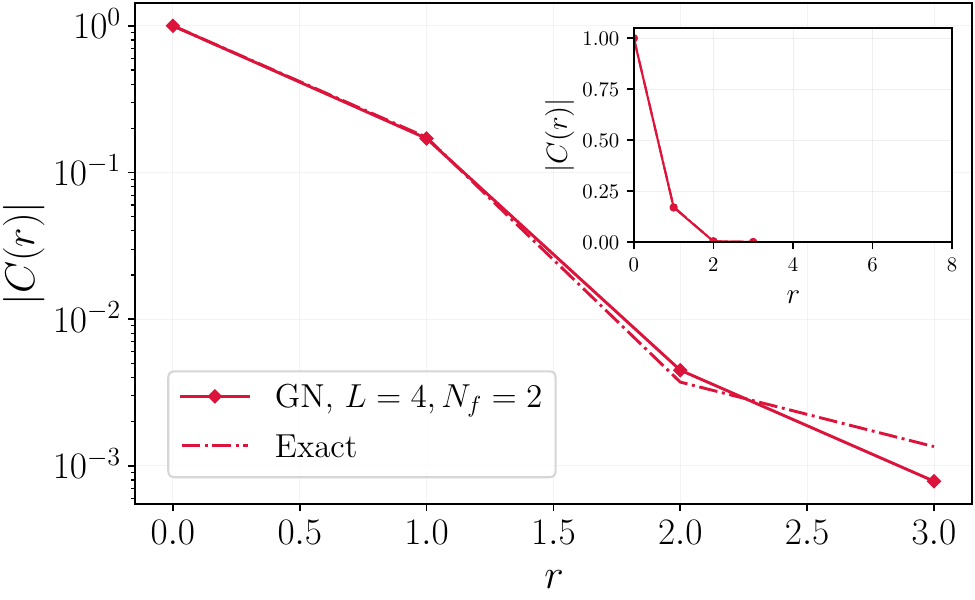}
        \caption{Gross--Neveu model}
        \label{fig:corr_AVQITE_gn}
    \end{subfigure}

    \caption{
    Normalized correlator defined by Eq.~\eqref{eq:Cr_definition} computed using the AVQITE-prepared ground state $\vert \psi_{\text{AVQITE}}\rangle$ (solid markers), compared to exact results computed using $\vert \psi_{\text{Exact}}\rangle$ (dashed-dotted lines) for the same datasets as considered in Fig.~\ref{fig:AVQITE_results1}. 
    }
    \label{fig:corr_AVQITE}
\end{figure}

As an application of our ground state preparation, we compute an equal-time, connected, averaged two-point function
of the site operator at site $i$, $(\overline{\psi}\psi)_{i} = \mathcal{N}_i = \sum_{f}(n_{i,f,1}-n_{i,f,2})$,
which is the operator expressed in terms of the number operator corresponding to the top and bottom component of the Dirac spinor. Let us define the connected correlator between two lattice sites $i$ and $j$ as: 
\begin{equation}
C_{\mathrm{conn}}(i,j)
= 
\langle \psi_0 \vert \mathcal{N}_i \mathcal{N}_j \vert \psi_0\rangle
-
\langle \psi_0 \vert \mathcal{N}_i \vert \psi_0\rangle\,\langle \psi_0 \vert \mathcal{N}_j \vert \psi_0\rangle
= 
\langle \mathcal{N}_i \mathcal{N}_j\rangle-\langle \mathcal{N}_i\rangle\langle \mathcal{N}_j\rangle .
\end{equation}
A useful observable that probes the
mass gap/correlation length is the averaged correlator $C(r)$ defined as: 
\begin{align}
C(r) &= \frac{1}{L-r}\sum_{i=0}^{L-r-1} C_{\mathrm{conn}}(i,i+r), \nonumber \\
& = 
\frac{1}{L-r}\sum_{i=0}^{L-r-1}
\Big(
\langle \psi_0 \vert \mathcal{N}_i \mathcal{N}_{i+r} \vert \psi_0\rangle
-
\langle \psi_0 \vert \mathcal{N}_i \vert \psi_0\rangle\,\langle \psi_0 \vert \mathcal{N}_{i+r} \vert \psi_0\rangle
\Big).
\label{eq:Cr_definition}
\end{align}
We show the results for the absolute value of the normalized $C(r)$ in Fig.~\ref{fig:corr_AVQITE}. Since we have a gapped Hamiltonian for $m=0.5, g = 0.2$, it decays exponentially as
should be the case for a gapped theory. The correlation function agrees well to within three digits of the exact result, which is roughly the same order as the fidelity.

\section{\label{sec:resource_costs}Hamiltonian simulation using higher-order product formulas and QSVT}

In this section, we calculate the resources required for the Hamiltonian simulation of the interacting four-fermion models in 1+1-dimensions. The results are summarized in Table.~\ref{tab:table_costs}. 

\begin{table}[h]
\centering
\renewcommand{\arraystretch}{1.15}
\begin{tabular}{|c|c|}
\hline
\textbf{Method} & \textbf{Cost scaling}\\
\hline
Order-$p$ Trotter
& $\mathcal{O} \left(L^{2} N_f^{4}t^{1+\frac{1}{p}}\epsilon^{-\frac{1}{p}}\right)$ \\
\hline
QSVT
& $O \Big(LN_f^2 t \,(N_f+\log(LN_f^2)) + (N_f+\log(LN_f^2))\log(1/\epsilon)\Big).$\\
\hline
\end{tabular}
\caption{\label{tab:table_costs}Summary of asymptotic costs for order-$p$ product formula and QSVT approaches to Hamiltonian simulation. Here, $L$ is the number of lattice sites, $N_f$ denotes the number of fermion flavors, $t$ is the total simulation time, and $\epsilon$ denotes the error in approximating the exact unitary $\exp(-iHt)$.}
\end{table}

\subsection{Complexity using product formulas}

One popular method for Hamiltonian simulation is the product formula, also referred to as the `Suzuki-Trotter' or `Lie-Trotter-Suzuki' formula~\cite{Hatano:2005gh}. The basic idea of this approach is as follows. Suppose, we want to understand dynamics generated by a given Hamiltonian $H$ which is written as sum of different terms with coefficient $c_j$ and Pauli string $P_{j}$ as $H = \sum_{j=1}^{m} H_{j} = \sum_{j=1}^{m} c_{j} P_{j}$. We can split the exponential of the sum by product of exponentials corresponding to each term by incurring some error due to the non-commutativity of the terms. For example, the first-order product or Trotter formula is:
\begin{equation}
e^{-iHt} =  \Big(\prod_{j=1}^{m} e^{-iH_{j}t/r}\Big)^{r} + \underbrace{\mathcal{O}\Big(\sum_{j < k} \bigg\vert\bigg\vert [H_j, H_k] \bigg\vert\bigg\vert \frac{t^{2}}{r}}_\epsilon \Big).
\end{equation}
If Hamiltonian only consists of commuting terms, the Trotter accuracy $\epsilon$ is zero. Following the works of Refs.~\cite{Childs2019, Childs:2019hts, Zhuk:2023zeh}, it is known that given a local Hamiltonian with nearest-neighbor interactions, we need $O(n^2 t^2)$ gates using the first-order product formula for some fixed accuracy $\epsilon$. This can be improved to $O(n^{1+1/p} t^{1+ 1/p})$ where $n$ is the number of qubits, $t$ is the total evolution time, and $p$ is the order of the product formula used. This result is optimal and consistent with those obtained using the Lieb-Robinson bound~\cite{Haah:2018ekc}. For 
lattice Hamiltonians with a two-component Dirac spinor with $O(N_f)$ geometric locality and maximum Pauli weight, we are
not aware whether such tight bounds exist.

Once we have the Hamiltonian in the second-quantized form, we use the Jordan--Wigner mapping to express it in terms of Pauli strings of length $2N_fL$ where the maximum weight is $2N_{f}+2$. In addition to the Jordan--Wigner mapping, we have also tried using Bravyi--Kitaev mapping; however, the difference was not significant for the range of qubits considered. Then, we can use a graph-coloring algorithm to collect the Pauli strings into clusters consisting of commuting Pauli string. For the Gross--Neveu model and the Thirring model with $N_f=1$, we have $\Gamma = 3$ clusters, while for the Thirring model with $N_f > 1$, we have $\Gamma= 4$. Once we have these clusters, we can simultaneously diagonalize each cluster and find its unitary circuit~\cite{van_den_Berg_2020, Murairi:2022zdg, Asaduzzaman:2023wtd, Jha:2024vcw}. For each cluster, finding the diagonalizing unitary circuit amounts to finding $V$ such that $V^{\dagger}HV  = H_{d}$, where $H_{d}$ consists of Pauli strings with either $Z$ or $\mathbb{I}$ (diagonal).
The Trotter error comes from the terms in different clusters that do not commute with each other. The sum of spectral norms for each pair using the first-order product formula is bounded by $O(LN_f^2)$. The number of steps $n$ required to achieve error $\epsilon$ is ($j, k = 1, \dots, \Gamma$): 
\begin{align}
    n &\ge \frac{t^2}{2\epsilon} \sum_{1\le j<k\le \Gamma} \Big \vert  \Big \vert [H_j, H_k] \Big \vert \Big  \vert \nonumber \\
    &\ge \frac{t^2}{2\epsilon} \widetilde{\alpha},
\end{align}
where $\widetilde{\alpha} = \sum_{1\le j<k\le \Gamma} \Big \vert  \Big \vert [H_j, H_k] \Big \vert \Big  \vert$ is the commutator norm, which is $O(L N_f^{2})$. Hence, we find that for each first-order Trotter step, we need $O(L N_f^{2})$
Clifford+T-gates leading to overall complexity of 
$O(L^{2} N_f^{4} t^{2}/\epsilon)$. We collect the total Clifford+T-gate costs using the first-order Trotter approach in  Table~\ref{tab:cx_counts1} and \ref{tab:cx_counts2} in
Appendix~\ref{sec:cx+t_appendix}. 

Instead of using the first-order product formula, we can use the higher-order product formulas. We now estimate the costs of the Hamiltonian simulation based on higher-order (order $p$) product formulas. The general bound for the number of required Trotter steps $n_p$ is:
\begin{align}
    n_p &\ge \Gamma t \left( \frac{\widetilde{\alpha}_{p+1} \,t}{\epsilon} \right)^{1/p} \nonumber \\ 
    &\ge  \Gamma t \left( \frac{t}{\epsilon} \right)^{1/p} \widetilde{\alpha}^{\frac{p+1}{p}},
\end{align}
where $\widetilde{\alpha}_{p+1}$ is the $p+1$-nested commutator sum:
\begin{equation}
\widetilde{\alpha}_{p+1}
=
\sum_{j_1,\dots,j_{p+1}=1}^\Gamma
\Big \vert \Big \vert
\big[H_{j_{p+1}},\big[H_{j_p},\dots,\big[H_{j_2},H_{j_1}\big]\dots\big]\big]
\Big \vert \Big \vert \le 2^p \alpha^{p+1},
\end{equation}
the inequality follows from repeated use of $\vert \vert [A,B] \vert \vert \le 2 \vert \vert A \vert \vert \, \vert \vert B \vert \vert$, and $\vert \vert \cdot \vert \vert$ denotes the spectral norm. We need (in the worst-case) $n \ge O(L N_f^2 t^{1 + \frac{1}{p}}\epsilon^{-1/p})$ steps, leading to overall complexity of $O(L^2 N_f^4 \,t^{1+\frac{1}{p}} \epsilon^{-1/p})$
for order $p$ product formula. For local Hamiltonians, there exists a tighter result of $O(N_f L t)$ due to the Lieb-Robinson bound~\cite{Childs2019}. However, for our case of relativistic fermionic models with large $N_f$ mapped to qubits, it is not clear whether such a tight bound can be attained. We leave this question for future work. 

\subsection{Cost estimate based on qubitization/QSVT based simulation}

In addition to computing the complexity using the higher-order product formula discussed above, an alternative approach to Hamiltonian simulation is based on the idea of block encoding/qubitization and quantum singular value transformations~\cite{PhysRevLett.118.010501,Gilyen:2018khw,Low2019hamiltonian,Gilyen:2018khw,Dong:2021qau,Martyn:2021bcm, Berry:2024ghc}. We refer the reader to ~\cite{PRXQuantum.2.040203,Joven2026} for reviews. We are interested in the complexity of the cost of Hamiltonian simulation for large $L$ and $N_f$ for four-fermion models. 

\subsubsection{Cost of block-encodings}
In order to estimate the cost using quantum singular value transformations (QSVT)/ qubitization for the Hamiltonian simulation, the first step is to block-encode the Hamiltonian (Hermitian matrix) $H$, or more accurately, $H/\alpha$ where $\alpha$ is a rescaling factor that keeps the spectrum of $H/\alpha$ in the interval $[-1, 1]$. 
A unitary $U$ is said to be an $\varepsilon$-block-encoding of $\widetilde{H} = \frac{H}{\alpha}$ provided that: 
\begin{equation}
\Big \vert  \Big \vert
\widetilde{H} - \big(\bra{0^a}\otimes I_s\big)\, U \,\big(\ket{0^a}\otimes I_s\big)
\Big \vert  \Big \vert \le \frac{\varepsilon}{\alpha},
\end{equation}
where $\ket{0^a}$ denotes the ancilla qubits and the subscript $s$ denotes the system. 
i.e., the top-left block of $U$ approximates $H/\alpha$ within operator-norm error
$\varepsilon/\alpha$ as: 
\begin{equation}
U=
\begin{pmatrix}
H/\alpha & \, \,  * \\
*  & \,\,  *
\end{pmatrix}.
\end{equation}
The Hamiltonian for the Thirring and Gross--Neveu models in the limit of large $L$ and $N_f$ has $O(LN_f^2)$ terms (see Eq.~\eqref{eq:Pauli_terms_GN_thirring}), and can be written schematically as:  
\begin{equation}
    \widetilde{H} = \frac{1}{\alpha
    } \sum_{j=1}^{\gamma} \alpha_j P_{j},
\end{equation}
where $\gamma = 5N_f L + 2 L N_f^2 - 4N_f$ for the GN model and  $\gamma = 3N_f L + 4 L N_f^2 - 4N_f$
for the Thirring model. Each $P_j$ is a Pauli string of length $2N_fL$ and has maximum weight $2N_f +2$. The first step is to prepare the index register so that: 
\begin{equation}
\text{PREP} (\ket{0}^{\otimes \log(O(LN_f^2))}) = \ket{A} = \sum_{j=1}^{O(LN_f^2)} \sqrt{\frac{\alpha_j}{\alpha}} \ket{j},
\label{eq:PREP_ket00}
\end{equation}
which prepares the coefficient, and then use another unitary 
$\text{SELECT}$, which applies the controlled Pauli operator as:
\begin{equation}
 \text{SELECT}(H) = \sum_{j=1}^{O(LN_f^2)}\ket{j}\bra{j} \otimes P_j.   
\end{equation}
Then the (normalized) Hamiltonian $\widetilde H:=H/\alpha$ is block-encoded as: 
\begin{equation}
U :=
\Big(\mathrm{PREP}^\dagger \otimes \mathbb{I}\Big) \cdot
\mathrm{SELECT}(H) \cdot
\Big(\mathrm{PREP}\otimes \mathbb{I} \Big).
\end{equation}
Similarly to $U$, we can also do this via the standard qubitization walk operator, $W$, which is more suitable for input to processing with 
QSVT. One query to $W$ consists of one call to $\mathrm{SELECT}$ and two calls to $\mathrm{PREP}$. Each QSVT step uses one query to the $\varepsilon$-block-encoding in the form of $W$. The 
explicit form of the qubitization walk operator is $W = \Big[2( \text{PREP}\ket{0}\bra{0} \text{PREP}^{\dagger}) - \mathbb{I} \Big] \cdot \,\text{SELECT}$ (for example, see 
Corollary 8 of Ref.~\cite{Motlagh:2023oqc}). 

Using the definition of $\mathrm{PREP}$ and 
$\text{SELECT}$, we now estimate the gate costs for one query to $W$. The cost of 
$\mathrm{PREP}$ depends on the coefficients of the Hamiltonian $H$. If all the $O(LN_f^2)$ terms of $H$ are the same, i.e., $\alpha_j = c$ for all $j$, then we can apply $m = O(\log(LN_f^2))$ Hadamard gates to obtain the uniform superposition: 
\begin{equation}
  \frac{1}{\sqrt{2^m}}\sum_{j=0}^{2^m-1}\ket{j},  
\end{equation}
leading to a total Clifford+T-gate cost of $O(1)$
gates. The other extreme case is where all terms are different, leading to a cost of $O(LN_f^{2})$. For us, the intermediate case is of interest
for the Hamiltonian with $K=3$ groups with a definite value of the kinetic mass and a four-fermion term with $K \ll O(LN_f^2)$. In such a case, $|\alpha_j|$ is constant within each group, and we can prepare a superposition over each group together and then combine them using some controlled operation. This leads to a cost of $O(\log K + \log LN_f^2)$, which gives a cost of $O(\log LN_f^2)$ for the PREP circuit
since $K \ll O(LN_f^2)$. Next, we calculate the cost for the $\mathrm{SELECT}$ circuit. 
For this, we require an index register of size $m=\lceil\log(LN_f^2)\rceil$ to compute a 1-bit equality flag and apply a single flag-controlled Pauli string. For our Hamiltonian, the maximum Pauli weight is $2N_f + 2$, so one SELECT call costs $O(N_f+\log(LN_f^2)\big)$~\cite{Zindorf:2024qat}, and using this we find that the leading cost of one query to $W$ is $O(N_f+\log(LN_f^2)\big)$. 

\subsubsection{Overall cost of the simulation}

With an $\varepsilon$-block-encoding of $H/\alpha$ as $W$, QSVT can implement an $\varepsilon$-block-encoding of $e^{-iHt} = e^{-i (H/\alpha) (\alpha t)}$ by $d$ queries to $W$, where $d$ is the degree of the polynomial $P_d$ that $\epsilon$-approximates
$e^{-i\alpha t x}$ uniformly on $x\in[-1,1]$, i.e.,
\[
\max_{x\in[-1,1]}\,\Big \vert P_d(x) - e^{-i\alpha t x} \Big \vert \le\;\epsilon.
\]
For a target error $\epsilon$, a standard estimate for required polynomial degree, $d$ can be obtained from Corollary 6 of Ref.~\cite{Low:2016znh} as:
\begin{equation}
   d = O(\alpha t + \log(1/\epsilon)\big), 
\end{equation}
where $\alpha$ is the unitary-invariant spectral norm of $H$ denoted $\vert \vert H \vert \vert$, $t$ is the simulation time, and $\epsilon$ is the accuracy.
For the four-fermion models, assuming that the Hamiltonian parameters ($m$, $g$) are all $O(1)$, we have $\alpha = O(LN_f^2)$ in the limit of large $L$ and $N_f$. This leads to a QSVT polynomial degree of complexity of $O(L N_f^2 t + \log(1/\epsilon))$, which combined with the cost of each query to $W$ gives the total gate complexity of 
\[
O \Big(LN_f^2 t \,(N_f+\log(LN_f^2)) + (N_f+\log(LN_f^2))\log(1/\epsilon)\Big).
\]

\subsection{Comparison between higher-order Trotter and QSVT}

\begin{figure}[h]
    \centering

    \begin{subfigure}[t]{0.49\linewidth}
        \centering
        \includegraphics[width=\linewidth]{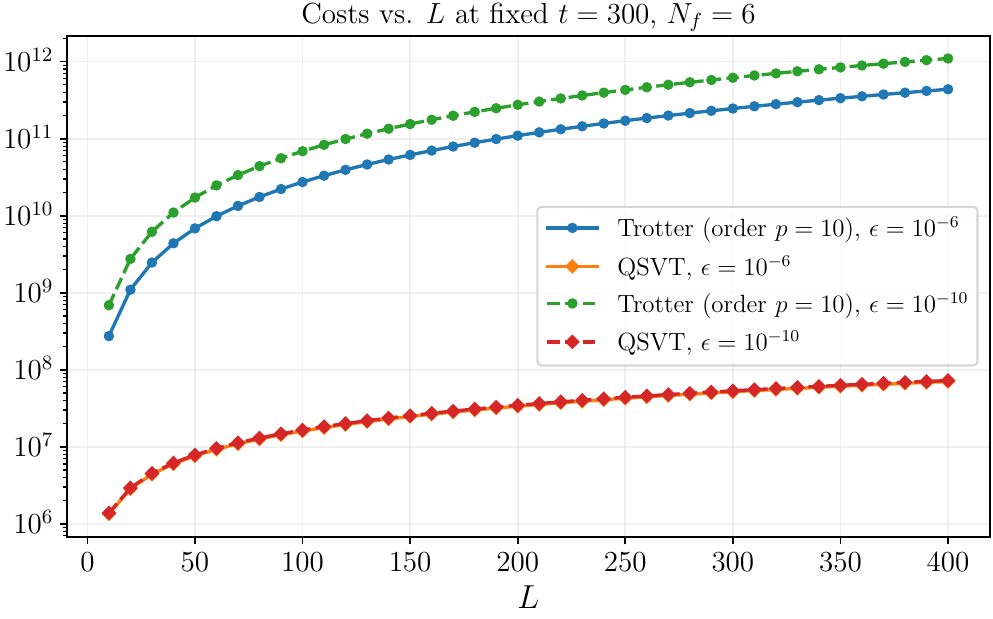}
        \caption{Cost vs.\ system size $L$ for $p=10$ comparing the product formula and QSVT at fixed $N_f=6$ and $t=300$, shown for $\epsilon = 10^{-6},10^{-10}$.}
        \label{fig:costs_L_comp}
    \end{subfigure}
    \hfill
    \begin{subfigure}[t]{0.49\linewidth}
        \centering
        \includegraphics[width=\linewidth]{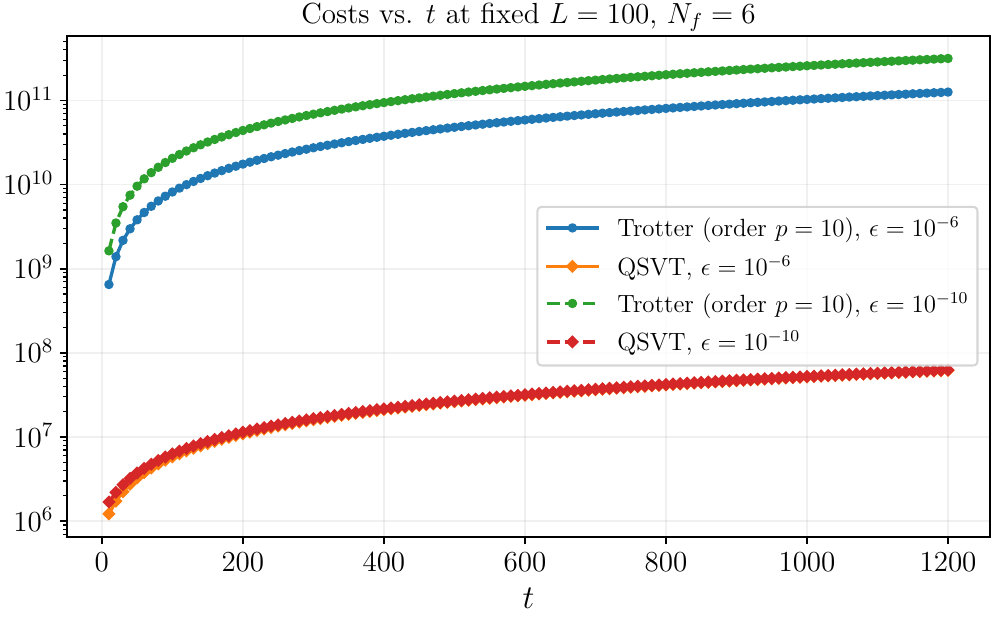}
        \caption{Cost vs.\ simulation time $t$ for $p=10$ comparing the product formula and QSVT at fixed $N_f=6$ and $L=100$, shown for $\epsilon = 10^{-6},10^{-10}$.}
        \label{fig:costs_t_comp}
    \end{subfigure}

    \caption{Comparison of asymptotic simulation costs for a $p$-th order product-formula (Trotter) method and QSVT.
    }
    \label{fig:costs_all_comp}
\end{figure}

In the previous subsections, we provided the complexity using higher-order Trotter and QSVT methods. We now show the comparison of the costs under various assumptions. Let us first start with dependence of costs on $L$ with other parameters fixed. In particular, 
we plot the following: 
\begin{align}
\mathrm{Cost}_{\rm PF}(L,t,N_f,\epsilon,p) &= L^{2} N_f^{4}\, t^{1+1/p}\, \epsilon^{-1/p},\\
\mathrm{Cost}_{\rm QSVT}(L,t,N_f,\epsilon) &= L N_f^{2}\Big[t\,g(L,N_f)+g(L,N_f)\log(1/\epsilon)\Big],
\end{align}
with $g(L,N_f)=N_f+\log(LN_f^{2})$. 
For the plots we set the constant factors, $C_{\rm PF}=C_{\rm QSVT}=1$, which we note is not rigorous. The result for cost dependence on $L$ with all other parameters fixed
is shown in Fig.~\ref{fig:costs_L_comp}. 
The result for cost dependence on simulation time $t$ for all other parameters fixed is shown in Fig.~\ref{fig:costs_t_comp}. 
QSVT has a clear advantage over product formulas, and this is related the lattice/qubit structure due to large $N_f$ shown in Fig.~\ref{fig:cartoon1}, which is more complicated than the nearest-neighbor (local) spin model simulations where QSVT and Trotter perform equally well at sufficiently large $p$~\cite{Childs2019}. 

\section{\label{sec:DLA_class}Dynamical Lie algebra of four-fermion models for general $N_f$}

We now discuss the classification of the dynamical Lie algebra (DLA) associated with the four-fermion models considered in this paper.  The structure and size of the DLA are essential for understanding quantum control~\cite{DAlessandro2021}, useful for accelerating the path to quantum technologies, training hardness~\cite{Ragone:2023qbn}, and fast-forwarding of certain Hamiltonians~\cite{Kokcu:2021ctj, Alsheikh:2025gbh}. We construct the DLA $\mathfrak{g}(H)$ generated by the Pauli strings in the Hamiltonian $H$ for general $N_f$. We give some details about DLA in Appendix~\ref{sec:dla_appendix} and refer the readers to Refs.~\cite{DAlessandro2021, Schirmer:2002kuu, Khaneja:2000stb,Goh:2023kcm,Wiersema:2023txu, Kazi:2025qpb} for more details.

Our main result is that the DLA for both four-fermion models considered in this work belong to Class B3 as per Ref.~\cite{Aguilar:2024dnz} mentioned in Table~\ref{tab:DLA_all}. The generated Lie algebra does not act irreducibly on the full Hilbert space of size $2^{2N_fL}$, but instead decomposes it into many disconnected symmetry sectors, which are the `superselection sectors'. For the GN model, we have $2N_f$ such sectors, while for the Thirring model, we have $N_f +1$.  
Therefore, the DLA of each irreducible sector is $\mathfrak{su}(2^{2N_fL}/2^{2N_f}) = \mathfrak{su}(2^{2N_fL-2N_f})$
for the Gross--Neveu model, and  
$\mathfrak{su}(2^{2N_fL}/2^{N_f+1}) = \mathfrak{su}(2^{2N_fL-N_f-1})$
for the Thirring model. We present these results in  
Table~\ref{tab:DLA_all}.

The structure of the DLA for a given Hamiltonian (generating set) does not depend on the number of qubits of the system; thus, it suffices to consider a few simple test cases numerically and then predict the isomorphism class. From the work of Refs.~\cite{Wiersema:2023txu,Kokcu:2024hla}, it is known that any generating set consisting purely of quadratic fermionic operators can be mapped to the DLA of free fermions, which is polynomial in the number of qubits $n$. However, for both models considered here, the quartic terms are present at non-zero value of the four-fermion coupling $g$, and generate exponential DLAs. We present the general case of $L$ lattice sites with $N_f$ flavors of fermions for both models 
in Table~\ref{tab:DLA_all}. The DLA determines the set of states that can be reached during the evolution of the system and is related to its controllability~\cite{DAlessandro2021}, and its trainability and presence of barren plateaus~\cite{Ragone:2023qbn}. It also has connections to the time-energy uncertainty principle~\cite{Atia:2016sax} and fast-forwarding of certain Hamiltonians~\cite{Kokcu:2021ctj, Gu:2021hyo, Alsheikh:2025gbh}. 

\begin{table}[h]
  \centering
  \setlength{\tabcolsep}{3pt} 
  \renewcommand{\arraystretch}{1.5}
  \begin{tabular}{c|c}
        Thirring       & Gross--Neveu \\ 
    \hline  \hline 
     $\mathfrak{g} = \oplus_{i=1}^{2^{N_f+1}} 
    \mathfrak{su}(2^{2N_fL - N_f - 1})$ &  
    $\mathfrak{g} = \oplus_{i=1}^{2^{2N_f}} 
    \mathfrak{su}(2^{2N_fL - 2N_f})$
  \end{tabular}
  \caption{Structure of the dynamical Lie algebra for both models for general $N_f$ where $n = 2N_fL$.
  For both models with any $N_f$, the DLA belongs to Class B3, according to the classification of Ref.~\cite{Aguilar:2024dnz}. For $N_f=1$, the DLAs are the same, since the models are equivalent due to the Fierz identity.}
  \label{tab:DLA_all}
\end{table}

The exponential dimension of the DLAs we have found for any $N_f$ 1+1-dimensional four-fermion models is not surprising. If we set the four-fermi interaction term coupling $g = 0$, we obtain polynomial DLA, since the remaining terms are all quadratic in the fermionic operators~\cite{Wiersema:2023txu,Kokcu:2024hla}. For $N_f=1$, we find that the DLA is same for both models, which correctly captures the equivalence between these models due to the Fierz identity. Our results for the exponential DLA size also relate to future simulations of some condensed matter systems. In particular, the four-fermion models we have considered can be mapped to 1d Fermi--Hubbard model at half-filling~\cite{Melzer1995}. However, taking that limit does not change the exponential increase of the DLA dimension. Therefore, our results also signal that such cases are likely to be affected by problems due to the vanishing variance of the cost function leading to the barren plateau problem~\cite{Ragone:2023qbn}.

\section{\label{sec:summary_fd}Summary and future directions}

We have provided the quantum-computational framework to simulate the 1+1-dimensional Thirring and Gross--Neveu models with arbitrary number of fermionic flavors, $N_f$. Using our formulation, we then constructed the ground state of both the Thirring and Gross--Neveu models for $N_f = 1, 2, 3, 4$ up to 20 qubits using AVQITE methods, and found excellent agreement with the exact results. We then considered the problem of complexity of Hamiltonian simulation using both higher-order product formulas and block encoding/QSVT methods and found that the required resources scale polynomially in both system size and number of flavors and linear in simulation time, which suggests that our quantum simulation is efficient. Finally, we identified the dynamical Lie algebra (DLA) of both models and found that they belong to the same isomorphism class of Ref.~\cite{Aguilar:2024dnz}. 

Our results for both ground state preparation and time evolution provide concrete evidence that quantum computing methods are efficient for understanding non-perturbative aspects of four-fermionic models through real-time evolution and for exploring interesting questions that are of interest to high-energy theorists at finite and large $N_f$. In the coming years, we believe that this work will be extended to a full-scale hardware implementation on near-term and fault-tolerant devices, paving the way to understanding relativistic fermionic field theories with properties such as asymptotic freedom similar to four-dimensional QCD. In addition, it would be interesting to apply other quantum algorithms for state preparation and see how they compare with the results presented here using AVQITE. Future computations of these relativistic field theories on quantum hardware in 1+1-dimensions and eventually in higher dimensions will help us probe and better understand various aspects of chiral symmetry breaking, flavor physics, and eventually lead to complete understanding of real-time and non-perturbative processes in QCD. 

\section*{Data Availability Statement}

The data used in this paper can be obtained from
Ref.~\cite{Dataset2026}. 

\section*{Acknowledgements}

R.G.J. would like to thank Simon Catterall, George Siopsis, Robert Edwards, Goksu Can Toga, Omar Alsheikh, Jack Araz, and Bharath Sambasivam for discussions. R.G.J., who planned and led the project,  B.N.B., A.F.K., and Y.L. were supported by the U.S. Department of Energy, Office of Science, Advanced Scientific Computing Research, under contract number DE-SC0025384. The AVQITE simulations (J.C.G.) were supported by the U.S. Department of Energy, Office of Science, Office of Advanced Scientific Computing Research under Award Number DE-SC0025623. We acknowledge the computing resources provided by North Carolina State University High Performance Computing Services Core Facility (RRID:SCR\_022168).

\bibliographystyle{utphys}

\bibliography{refs.bib}

\clearpage
\appendix
\renewcommand\thefigure{A\arabic{figure}}  
\renewcommand\thetable{A\Roman{table}}  
\setcounter{figure}{0}
\setcounter{table}{0}

\section{\label{sec:sampleH_appendix}Example construction of qubit Hamiltonian}
Let us start by writing the Hamiltonian for $L=2, N_f = 2$ models to clarify the construction. For $N_f = L=2$, we have an 8-qubit Hamiltonian composed as:
\begin{align}
    H = H_k + H_m + H_{\text{FF}}.
\end{align}
Using the Jordan--Wigner (JW) transformation, we see that four pairs of $c$, i.e., $\{c_0, c^{\dagger}_0, \cdots, c_3, c^{\dagger}_3\}$ corresponding to four qubits are on the first site and $\{c_4, c^{\dagger}_4, \cdots, c_7, c^{\dagger}_7\}$ are on the second site. Recall that we are using open boundary conditions. 
The mass term of the Hamiltonian is given by: 
\begin{align}
    H_m &= m \sum_{i=1}^{L} \overline{\psi_i} \psi_i = m \sum_{i=1}^{L} (c^{\dagger}_{u, i}c_{u, i}  - c^{\dagger}_{d, i}c_{d, i}) \nonumber \\
    &= m(c^{\dagger}_0 c_0 + c^{\dagger}_2 c_2 + c^{\dagger}_5 c_5 + c^{\dagger}_7 c_7 - c^{\dagger}_1 c_1 - c^{\dagger}_3 c_3 - c^{\dagger}_4 c_4 - c^{\dagger}_6 c_6),
\end{align}
and leads to a total of $2N_fL$ terms. The interaction term is
\begin{align}
    H_{\text{FF}} &= -g (c^{\dagger}_0 c_{0}c^{\dagger}_0 c_{0}  + c^{\dagger}_0 c_{0}c^{\dagger}_1 c_{1} + c^{\dagger}_0 c_{0}c^{\dagger}_2 c_{2} + c^{\dagger}_0 c_{0}c^{\dagger}_3 c_{3} + 
    c_{1}c^{\dagger}_1 c_{1}
    c^{\dagger}_1 + c^{\dagger}_1 c_{1}c^{\dagger}_2 c_{2} + c^{\dagger}_1 c_{1}c^{\dagger}_3 c_{3} \nonumber \\ 
    &+ 
    c^{\dagger}_2 c_{2}c^{\dagger}_2 c_{2}
    + c^{\dagger}_2 c_{2}c^{\dagger}_3 c_{3} + 
    c^{\dagger}_3 c_{3}c^{\dagger}_3 c_{3} +
    [\text{same terms with~} c_{i} \to i+2N_f = i+4] + \text{H.c.},
\end{align}
where $i \to i + 4 $ means that we have the same ten terms but on the second site. For a general $N_f$, this will be $i \to i + 2N_f$ instead. This leads to a total of 20 terms. 
The Hermitian conjugate gives another 12 terms, adding to 32 terms and giving a total of $4LN_f^2$
terms. It is straightforward to show that for the Thirring model, we have another $4LN_f^2$ terms in $H_{\text{FF}}$.
Now, we come to the kinetic terms given by:
\begin{equation}
 H_k = -\frac{i}{2}\left(c^{\dagger}_{0}c_5  + c^{\dagger}_{1}c_4 + c^{\dagger}_{2}c_7 + c^{\dagger}_{3}c_6 - c^{\dagger}_{4}c_1 - c^{\dagger}_{5}c_0 - c^{\dagger}_{6}c_3 -
 c^{\dagger}_{7}c_2
 \right),  
\end{equation}
giving a total of $4N_f(L-1)$ terms. Note that the last four are Hermitian conjugates of the first four terms.
Adding these leads to a total number of terms in the Hamiltonian given by: 
\begin{align}
\text{GN:} \quad & 
6N_f L + 4 L N_f^2 - 4N_f, \nonumber \\
\text{Thirring:} \quad & 
6N_f L + 8 L N_f^2 - 4N_f.
\label{eq:terms_GN_thirring}
\end{align}
Once the Hamiltonian is expressed in terms of fermionic operators, we can then rewrite this in terms of Pauli strings using the JW transformation with a maximum weight of $2N_f +2$ for both models, which is independent of the number of spatial sites. The number of Pauli terms is:
\begin{align}
\text{GN:} \quad & 
5N_f L + 2 L N_f^2 - 4N_f, \nonumber \\
\text{Thirring:} \quad & 
3N_f L + 4 L N_f^2 - 4N_f,
\label{eq:Pauli_terms_GN_thirring}
\end{align}
which scales as $O(LN_f^2)$ in the large $N_f$ and the thermodynamic limit (large $L$). 

\section{\label{sec:qite_avqite_appendix}Additional details about QITE and AVQITE}

In this Appendix, we provide further details on the ground-state preparation method employed in this work, namely AVQITE~\cite{Gomes:2021ckn,Getelina:2023nwx,Getelina:2023yrf}, which is an adaptive-variational extension of QITE~\cite{Motta:2020qite}.

This section is divided into two parts. In the first part, we present a bound analysis on general imaginary-time evolution methods, while providing an expression for estimating the evolution time necessary to obtain a state within the desired precision. In the second part, we focus on the derivation of the objective function for AVQITE, starting from the McLachlan's variational principle~\cite{McLachlan:1964,Yuan:2018jdl}. Furthermore, we provide additional simulation details to ensure the reproducibility of our results.

\subsection{Convergence of QITE with non-zero overlap}
\counterwithin{figure}{section}

Consider a Hamiltonian $H$ that has a ground state $|g\rangle$ with energy $E_0$ and 
excited states $\{|k\rangle\}_{k\ge 1}$ with energies $E_k \ge E_0 + \Delta$, where $\Delta > 0$ is the spectral gap defined as $E_1 - E_0$. For an initial state $|\Psi_0\rangle$ that has a finite overlap $\delta$ with the target ground state, i.e., $ \delta \equiv |\langle \Psi_0 | g \rangle|\neq 0$, one can write the following expansion: 
\[
|\Psi_0\rangle = \delta |g\rangle + \sum_{k\ge 1} c_k |k\rangle. 
\]
Applying the imaginary time evolution operator to the above expression gives
\[
|\phi_\tau\rangle = e^{-\tau (H-E_0)} |\Psi_0\rangle 
= \delta |g\rangle + \sum_{k\ge 1} c_k e^{-\tau (E_k-E_0)} |k\rangle .
\]
Now, let us define the quantity 
\begin{equation}
  r_\tau \equiv \sum_{k\ge 1} |c_k|^2 e^{-2\tau (E_k-E_0)}.
\label{eq:r_tau1}
\end{equation}
Note that $e^{-2\tau(E_k - E_0)} \le e^{-2\tau\Delta}$ for $k\ge 1$, because
every excited states is at least one gap above the ground state. Hence, we can bound Eq.~\eqref{eq:r_tau1} as: 

\begin{equation}
    r_\tau \le (1-\delta^2) e^{-2\Delta \tau}. 
\end{equation}
The normalized state is
\begin{equation}
 \vert\Psi_\tau\rangle = \frac{\vert\phi_\tau\rangle}{\vert\vert \phi_\tau \vert\vert}, 
\qquad  \vert \vert \phi_\tau \vert \vert^2 = \delta^2 + r_\tau .  
\end{equation}
Thus, the fidelity with the true ground state after evolving for time $\tau$ is:
\begin{equation}
F(\tau) = \vert \langle g \vert \Psi_{\tau} \rangle \vert^2
=\frac{\delta^2}{\delta^2+r_\tau}
=\frac{1}{1+r_\tau/\delta^2}.
\end{equation}
Since $(1+x)^{-1}\ge 1-x$ for all $x\ge 0$, we obtain the lower bound
\begin{equation}
F(\tau) \ge 1-\frac{r_\tau}{\delta^2}.
\end{equation}
Using $r_\tau\le (1-\delta^2)e^{-2\Delta\tau}$, this yields
\begin{equation}
F(\tau) \ge 1- \Big(\frac{1}{\delta^2} - 1 \Big) e^{-2\Delta\tau}.
\label{eq:fidelity-bound}
\end{equation}
To attain $\epsilon$ infidelity, $F(\tau) \ge 1-\epsilon$, we need to evolve to imaginary time $\tau$ given by:
\begin{equation}
\tau \ge  \frac{1}{2\Delta} \log \left(\frac{\frac{1}{\delta^2}-1}{\epsilon}\right).    
\end{equation}

\subsection{AVQITE optimization procedure}

In this subsection, we give the basic formalism of AVQITE approach used in the paper. We refer the reader to Refs.~\cite{Yuan:2018jdl} for additional details. Let us start from the real-time Schr\"odinger evolution equation: 
\begin{align}
\partial_t \ket{\psi(t)} = -\,i \hat H \ket{\psi(t)} .
\end{align}
Performing a Wick rotation $t \to -i\tau$, we obtain $\partial_\tau \ket{\psi(\tau)} = -\hat H \ket{\psi(\tau)}$ which is the 
imaginary-time Schr\"odinger equation. The solution is
\begin{align}
\ket{\psi(\tau)} = e^{-\tau \hat H}\ket{\psi(0)} ,
\end{align}
which projects onto the ground state as 
$\tau \to \infty$ assuming nonzero ground-state overlap $\delta^2$. We will use AVQITE to approximate this solution. One of the problems associated with this approach of finding ground states is that the operator $\exp(-\tau \hat{H})$ is not unitary. AVQITE aims to follow the (normalized) imaginary-time flow: 
\begin{align}
\partial_\tau \ket{\psi(\tau)}
= -\bigl(\hat H - \langle \hat H\rangle_\tau\bigr)\ket{\psi(\tau)},
\qquad
\langle \hat H\rangle_\tau \equiv \bra{\psi(\tau)}\hat H\ket{\psi(\tau)}.
\end{align}
Instead of representing the non-unitary operator $e^{-\tau \hat H}$ directly, AVQITE restricts to a unitary variational family
\begin{align}
\ket{\psi(\boldsymbol\theta)} = U(\boldsymbol\theta)\ket{\psi(0)},
\qquad
U(\boldsymbol\theta)=\prod_{k} e^{-i\theta_k A_k},
\end{align}
where each generator $A_k$ is Hermitian and selected from a predetermined pool to find the closest unitary to $e^{-\tau \hat H}$. To understand the cost function used in AVQITE, we start by recasting the flow equation in terms of density matrices. We can write the density matrix, $\widetilde \rho = \ket{\psi}\bra{\psi}$, where $\widetilde {\rho}$ denotes the unnormalized density matrix. Then
\begin{align}
\partial_\tau \widetilde \rho
&= \ket{\partial_\tau \psi}\bra{\psi} + \ket{\psi}\bra{\partial_\tau \psi} \\
&= -\,\hat H \ket{\psi}\bra{\psi} - \ket{\psi}\bra{\psi}\,\hat H \\
&= -\,\{\hat H,\widetilde \rho\}.
\end{align}
To preserve normalization, we define
\begin{align}
\partial_\tau \ket{\psi} = -\bigl(\hat H-\langle \hat H\rangle\bigr)\ket{\psi},
\qquad
\langle \hat H\rangle=\bra{\psi}\hat H\ket{\psi}
= \mathrm{Tr}(\hat H\rho),
\end{align}
where \(\rho=\ket{\psi}\bra{\psi}\) is the normalized density matrix. Then
\begin{align}
\partial_\tau \rho
&= -(\hat H-\langle\hat H\rangle)\rho - \rho(\hat H-\langle\hat H\rangle) \label{eq:partial_tau_rho_equal_L}\\
&= -\{\hat H,\rho\} + 2\langle\hat H\rangle \rho.
\end{align}
If we now define the Liouvillian superoperator:
\begin{align}
\mathcal{L}(\rho) \equiv -\{\hat H,\rho\} + 2\langle\hat H\rangle \rho,
\label{eq:L_superop}
\end{align}
we find: 
\begin{align}
\partial_\tau \rho = \mathcal{L}(\rho).
\label{eq:flow_equation1}
\end{align}
In AVQITE simulations of Eq.~\eqref{eq:flow_equation1}, the variational state $\rho$ is a function of parameters, i.e., $\rho(\boldsymbol{\theta})$ but we will simply write it as $\rho$ and we have approximately $\partial_\tau \rho =
\sum_{\mu=1}^{N_\theta} (\partial_\mu \rho)\,\dot\theta_\mu$ using the shorthand $\partial_\mu \equiv \frac{\partial}{\partial \theta^\mu}$.
We can now define the residual, which measures the deviation from the actual imaginary flow, as: 
\begin{align}
R(\boldsymbol{\theta},\dot{\boldsymbol{\theta}})
= \sum_{\mu=1}^{N_\theta} \partial_\mu \rho(\boldsymbol{\theta})\,\dot\theta_\mu
- \mathcal{L}\bigl(\rho(\boldsymbol{\theta})\bigr).
\label{eq:R_cost_func_def}
\end{align}
The variational principle due to McLachlan aims to minimizes the quadratic cost: 

\begin{align}
L^2
= \Big \vert \Big \vert R(\boldsymbol{\theta},\dot{\boldsymbol{\theta}})\Big \vert \Big \vert_2^2,
\label{eq:L2_cost_func}
\end{align}
where $\vert \vert \cdot \vert \vert_2$ is the Schatten-2 norm (also known as the Frobenius norm) defined as $\vert \vert R \vert \vert_2^2 = \mathrm{Tr}(R^\dagger R)$. If we now expand Eq.~\eqref{eq:L2_cost_func} using the definition in Eq.~\eqref{eq:R_cost_func_def}, we obtain: 
\begin{align}
L^2
&= \mathrm{Tr}\Big[
\Bigl(\sum_\mu \partial_\mu\rho\,\dot\theta_\mu - \mathcal{L}(\rho)\Bigr)^\dagger
\Bigl(\sum_\nu \partial_\nu\rho\,\dot\theta_\nu - \mathcal{L}(\rho)\Bigr)
\Big] \nonumber  \\
&= \sum_{\mu\nu}\dot\theta_\mu\dot\theta_\nu\,
\mathrm{Tr}\bigl[(\partial_\mu\rho)^\dagger(\partial_\nu\rho)\bigr]
-\sum_\mu \dot\theta_\mu\,\mathrm{Tr}\bigl[(\partial_\mu\rho)^\dagger\mathcal{L}(\rho)\bigr] -\sum_\nu \dot\theta_\nu\,\mathrm{Tr}\bigl[\mathcal{L}(\rho)^\dagger(\partial_\nu\rho)\bigr]
+\mathrm{Tr} \bigl[\mathcal{L}(\rho)^\dagger\mathcal{L}(\rho)\bigr].
\label{eq:L2_expanded_form}
\end{align}
For real parameters $\dot\theta_\mu\in\mathbb{R}$, we can define: 
\begin{align}
M_{\mu\nu} &\equiv \mathrm{Tr}\Big[(\partial_\mu\rho)^\dagger(\partial_\nu\rho)\Big], \label{eq:M_munu_def}\\
V_\mu &\equiv \Re\, \mathrm{Tr}\Big[(\partial_\mu\rho)^\dagger\mathcal{L}(\rho)\Big], \label{eq:V_mu_def0} \\
C &\equiv \mathrm{Tr}\Big[\mathcal{L}(\rho)^\dagger\mathcal{L}(\rho)\Big]
= \vert \vert\mathcal{L}(\rho)\vert \vert_2^2.
\end{align}
Here $M_{\mu\nu}$ is an $N_{\boldsymbol{\theta}} \times N_{\boldsymbol{\theta}}$ real symmetric matrix also referred to as the quantum Fisher information matrix~\cite{Koczor:2019pvr, Gomes:2021ckn}. For a pure state, i.e., $\rho=\ket{\psi}\bra{\psi}$, we have
\begin{align}
\partial_\mu\rho
&= \ket{\partial_\mu\psi}\bra{\psi} + \ket{\psi}\bra{\partial_\mu\psi}.
\label{eq:partial_rho1}
\end{align}
Using Eq.~\eqref{eq:partial_rho1} in Eq.~\eqref{eq:M_munu_def} with $\mathrm{Tr}(\ket{a}\bra{b}) = \langle b \vert a \rangle $, we obtain: 
\begin{align}
M_{\mu\nu}
&= 2 \mathfrak{R} \Big[
\langle \partial_{\mu} \psi \vert \partial_{\nu} \psi \rangle
+ \langle \partial_{\mu} \psi \vert \psi \rangle\,\langle \psi \vert \partial_{\nu} \psi \rangle \Big],
\label{eq:M_mu_nu_def}
\end{align}
where we denote the parametrized pure state $\ket{\psi(\boldsymbol{\theta})}$ simply by $\ket{\psi}$. 
Now, consider Eq.~\eqref{eq:V_mu_def0}. We can express $V_\mu$ using Eq.~\eqref{eq:partial_tau_rho_equal_L} i.e., \(\mathcal{L}(\rho)=-(\hat H-\langle\hat H\rangle)\rho-\rho(\hat H-\langle\hat H\rangle)\) as:  
\begin{align}
V_{\mu} &= -2 \mathfrak{R} \Big[\langle \partial_{\mu} \psi \vert \hat H \vert \psi \rangle \Big],
\label{eq:V_mu_def}
\end{align}
where we have again used the shorthand $\ket{\psi} = \ket{\psi(\boldsymbol{\theta})}$. 
Now consider the last term of Eq.~\eqref{eq:L2_expanded_form}. Setting $A \equiv \hat H - \langle\hat H\rangle \mathbb{I}$, we can write Eq.~\eqref{eq:L_superop} as: 
\begin{align}
\mathcal{L}(\rho) = -\{A,\rho\}.
\end{align}
As \(\rho\) is Hermitian, we obtain:

\begin{align}
C &=\mathrm{Tr} \Big[\mathcal{L}(\rho)^2\Big] \nonumber \\ 
&=\mathrm{Tr}\Big[\{A,\rho\}^2\Big], \nonumber \\
& = 2\,\mathrm{Tr}(\rho^2 A^2) + 2\,\mathrm{Tr}(\rho A\rho A)
\label{eq:C_mu_def}
\end{align}
after expanding \(\{A,\rho\}^2=(A\rho+\rho A)^2\) and using cyclicity of the trace. For a pure state \(\rho=\ket{\psi}\bra{\psi}\), we have $\rho^2 = \rho$ and \(\rho A \rho = \langle A\rangle\rho = 0\).
Hence, $\mathrm{Tr}(\rho A\rho A) = 0$ and
\begin{align}
C = 2\langle A^2\rangle
= 2\bigl(\langle \hat H^2\rangle_{\boldsymbol{\theta}} - \langle\hat H\rangle_{\boldsymbol{\theta}}^2\bigr)
= 2\,\mathrm{Var}_{\boldsymbol{\theta}}(\hat H).
\label{eq:C_term}
\end{align}
Then, the objective cost function given by Eq.~\eqref{eq:L2_expanded_form} becomes: 
\begin{align}
L^2 = \sum_{\mu\nu} M_{\mu\nu}\dot\theta_\mu\dot\theta_\nu
-2\sum_\mu V_\mu \dot\theta_\mu + C.
\label{eq:L_dist_2}
\end{align}
where $M_{\mu\nu}$, $V_{\mu}$ and $C$ are defined in 
Eq.~\eqref{eq:M_mu_nu_def}, Eq.~\eqref{eq:V_mu_def}, and Eq.~\eqref{eq:C_term}, respectively. Differentiating  Eq.~\eqref{eq:L_dist_2} with respect to $\dot\theta_{\mu}$, we obtain:
\begin{align}
\frac{\partial L^2}{\partial \dot\theta_\mu}=0
\quad\Rightarrow\quad
\sum_\nu M_{\mu\nu}\dot\theta_\nu = V_\mu.
\label{eq:L_deriv_equation}
\end{align}
In vector form, we have $\dot{\boldsymbol{\theta}}^{\,*} = M^{-1} V$,
and plugging back into Eq.~\eqref{eq:L_deriv_equation}, we find the minimum/optimal value as: 
\begin{align}
L_{2,\min}^2
= C - V^{T} M^{-1} V.
\end{align}
In practical calculations, for a given time step, one should append operators to the ansatz whenever the measured McLachlan distance $L^2$ is below a certain threshold $L_{\text{cut}}^2$, which we set to $L_{\text{cut}}^2 = 10^{-2}$. We also limit the number of operators that can be added to the ansatz to five at each time step. If $L^2$ is above the threshold even after adding five operators, we simply move on to the next time step. The imaginary-time evolution proceeds until the maximum energy gradient $\max \vert V \vert$ is below an empirically determined threshold of $V_\textrm{cut}=10^{-4}$, which has been observed to consistently yield a final state with excellent fidelity with respect to the exact ground state for various spin models~\cite{Gomes:2021ckn,Getelina:2023nwx,Getelina:2023yrf}.

Note that inverting the matrix $M$ defined in Eq.~\eqref{eq:M_munu_def} may cause numerical problems due to its condition number being too large. To circumvent this issue, we apply a ridge regression (also called Tikhonov regularization), which consists of shifting the diagonal values of $M$ by a relatively small constant~\cite{Hoerl:1970ridge}. Here, for all of our calculations we have considered $M\rightarrow M+\lambda\mathbb{I}$ with $\lambda=10^{-6}$.

\begin{figure}
    \centering
    \includegraphics[width=\linewidth]{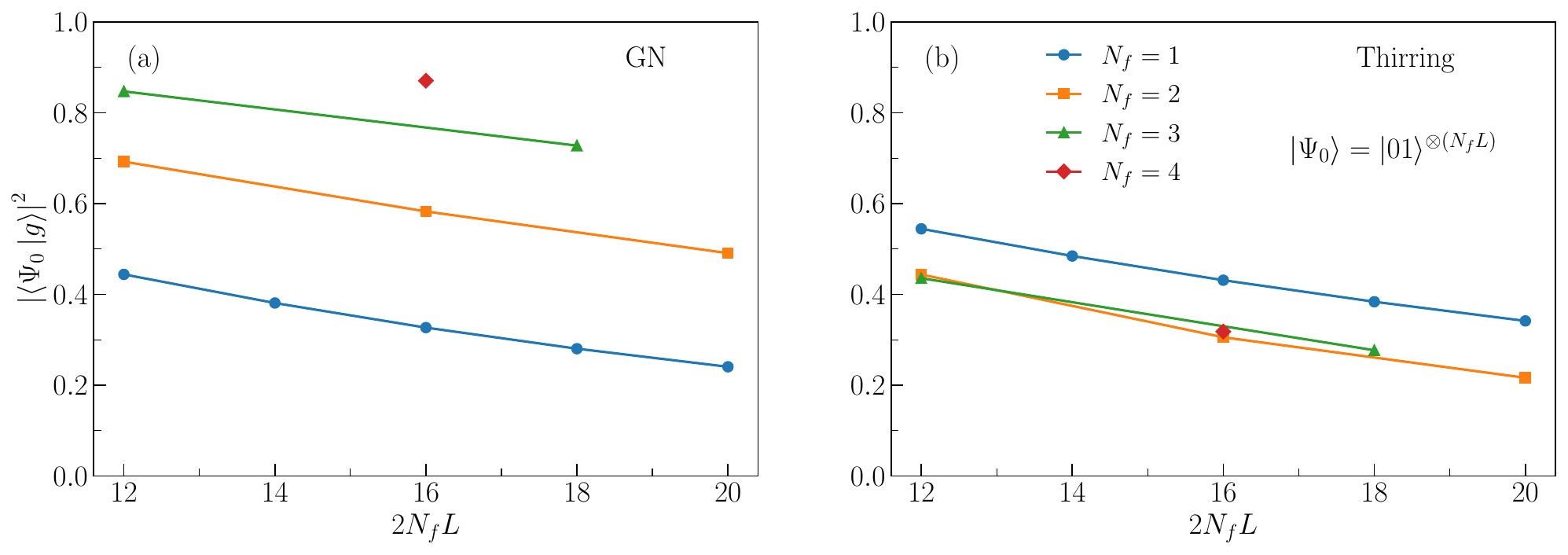}
    \caption{Overlap of the initial (N\'eel) state $\ket{\Psi_0}=\ket{01}^{\otimes (N_fL)}$ with respect to
    the exact ground state $\ket{g}$ vs. number of qubits for the (a) Gross-Neveu (GN) and (b) Thirring models.
    The data points include all systems considered in Table~\ref{tab:table_AVQITE}, with some additional points
    to better showcase the scaling trend.}
    \label{fig:overlap}
\end{figure}

Finally, as discussed in the previous subsection, having a finite overlap with the target ground state is essential for the success of any imaginary-time-evolution method. Fig.~\ref{fig:overlap} shows the initial state overlap as a function of the number of qubits for various system sizes and number of flavors for both the Thirring and Gross--Neveu models. For the Gross--Neveu model we see that the overlap improves with increasing $N_f$, while for the Thirring model it appears to be independent of $N_f > 1$. In both the cases, they seem to follow $O(\text{polylog}(n))$ reduction, which is better than the estimate of exponential decrease from naive counting.

\section{\label{sec:dla_appendix}Brief review of dynamical Lie algebras}

Let $\mathfrak{g}$ be a Lie algebra. A subalgebra $\mathfrak{s}$ of $\mathfrak{g}$ is defined as a subspace of $\mathfrak{g}$ that is closed
under the bracket operation, i.e., $[\mathfrak{s},\mathfrak{s}] \subseteq \mathfrak{s}$.
If a subalgebra $\mathfrak{s}$ satisfies $[\mathfrak{s},\mathfrak{g}] \subseteq \mathfrak{s}$, it is called an ideal of $\mathfrak{g}$.
A simple Lie algebra is a non-abelian Lie algebra $\mathfrak{g}$ (i.e., $[\mathfrak{g},\mathfrak{g}] \ne \{0\}$)
whose only ideals are $\{0\}$ and $\mathfrak{g}$. A Lie algebra is called semi-simple if it
can be written as a direct sum of simple (commuting) ideals.
For a semi-simple Lie algebra $\mathfrak{g}$, we can define its Cartan decomposition as a split
\[
  \mathfrak{g} = \mathfrak{k} \oplus \mathfrak{m}
\]
that satisfies: 
\[
  [\mathfrak{k},\mathfrak{k}] \subseteq \mathfrak{k}, \qquad
  [\mathfrak{m},\mathfrak{m}] \subseteq \mathfrak{k}, \qquad
  [\mathfrak{k},\mathfrak{m}] \subseteq \mathfrak{m}.
\]
In this paper, our focus is only on the dimension of $\mathfrak{g}$ rather than on finding the Cartan pair and the applications associated with it. In control theory~\cite{DAlessandro2021}, occurrence of barren plateaus~\cite{Ragone:2023qbn} and other problems, the size of the DLA, i.e., $\text{dim}(\mathfrak{g}(H))$ is often crucial. In a recent study~\cite{Wiersema:2023txu}, the DLAs of two-local spin models were classified, while the general case of a Pauli generator set was considered in Ref.~\cite{Aguilar:2024dnz}. Here, we determine the DLA for the fermionic Hamiltonians that correspond to four-fermion quantum field theories in the continuum limit. 

Let $\mathcal{G}=\{A,B,\ldots\}$ be a set of Pauli strings (Hermitian) on a finite-dimensional Hilbert space. The Lie bracket (also known as commutator) is defined as:
\begin{equation}
[A,B] = AB - BA.
\end{equation}
Since the commutator of Hermitian operators is anti-Hermitian, it is convenient to work in the real vector space of anti-Hermitian operators. The \emph{dynamical Lie algebra} generated by $\mathcal{G}$ is defined as
\begin{equation}
\mathfrak{g} = 
\mathrm{Lie}_{\mathbb{R}} \big(i\mathcal{G}\big)
= 
\mathrm{span}_{\mathbb{R}} \Big \langle iA,\, iB,\, \ldots \Big \rangle_{\mathrm{Lie}}
\label{eq:DLA_def}
\end{equation}
where $\langle\cdot\rangle_{\mathrm{Lie}}$ denotes the \emph{Lie closure}. The Lie closure $\langle i\mathcal{G}\rangle_{\mathrm{Lie}}$ is the smallest set that contains
$i\mathcal{G}$ and is closed under the Lie bracket, i.e., it is the set obtained by iterating commutators: 
\begin{equation}
X_1,X_2\in \langle i\mathcal{G}\rangle_{\mathrm{Lie}}
\quad\Rightarrow\quad
[X_1,X_2]\in \langle i\mathcal{G}\rangle_{\mathrm{Lie}},
\end{equation}
until no new elements are added.
Let us define such a basis of \emph{Hermitian} elements as $\{L_\mu\}_{\mu=1}^{\dim(\mathfrak{g})}$ so that:
\begin{equation}
\mathfrak{g} = 
\mathrm{span}_{\mathbb{R}}\{iL_\mu\}_{\mu=1}^{\dim(\mathfrak{g})},
\qquad
(L_\mu)^\dagger = L_\mu.
\label{eq:DLA_basis}
\end{equation}
Here $\dim(\mathfrak{g})$ is the dimension of the dynamical Lie algebra as a real vector space,
and $\{L_\mu\}$ is a set of linearly independent generators whose span determines $\mathfrak{g}$. For our case, the generating set $\mathcal{G}$ is simply the individual Pauli strings of the Hamiltonian $H$.

\section{\label{sec:cx+t_appendix}Gate counts using first-order product formula}

We provide the gate counts for the Hamiltonian simulation of the four fermion models using first-order product formulas in Table~\ref{tab:cx_counts1} and 
Table~\ref{tab:cx_counts2}. 

\renewcommand{\arraystretch}{1.1} 
\begin{table}[h]
\centering
\begin{tabular}{|c|c|c|c|c|}
\hline
\hline
$L$ & \# of $c, c^{\dagger}$ terms (Eq.~\eqref{eq:terms_GN_thirring}) & \# of Pauli strings (Eq.~\eqref{eq:Pauli_terms_GN_thirring}) & CX & Clifford+T\\
\hline
10  & 272  & 173 & 624  & 1400 \\
\hline 
20  & 552  & 353 & 1304 & 2940 \\
\hline 
40  & 1112 & 713 & 2664 & 6020\\
\hline 
100 & 2792 & 1793 & 6744  & 15260\\
\hline
\hline
\end{tabular}

\vspace{4mm}

\begin{tabular}{|c|c|c|c|c|}
\hline
\hline
$L$ & \# of $c, c^{\dagger}$ terms (Eq.~\eqref{eq:terms_GN_thirring}) & \# of Pauli strings (Eq.~\eqref{eq:Pauli_terms_GN_thirring}) & CX & Clifford+T \\
\hline
10  & 432  & 213 & 704 & 1960  \\
\hline 
20  & 872  & 433 & 1464 & 3460 \\
\hline 
40  & 1752 & 873 & 2984 & 8260 \\
\hline 
100 & 4392 & 2193 & 7544 & 20860  \\
\hline
\end{tabular}
\caption{Gate counts for varying lattice size \(L\) with fixed flavor number \(N_f = 2\) for GN model (top) and Thirring model (bottom) for one first-order Trotter step.}
\label{tab:cx_counts1}
\end{table}

\renewcommand{\arraystretch}{1.1} 
\begin{table}[h]
\centering
\begin{tabular}{|c|c|c|c|c|}
\hline
\hline
$N_f$ & \# of $c, c^{\dagger}$ terms (Eq.~\eqref{eq:terms_GN_thirring}) & \# of Pauli strings (Eq.~\eqref{eq:Pauli_terms_GN_thirring}) & CX & Clifford+T \\ 
\hline
2  & 272  & 173 & 624 & 1400 \\
\hline 
4  & 864  & 505 & 1604 & 3291   \\
\hline 
6  & 1776  & 997 & 2994 & 5538  \\
\hline 
8  & 3008  & 1649 & 4902 & 8348  \\
\hline
\hline
\end{tabular}

\vspace{4mm}

\begin{tabular}{|c|c|c|c|c|c|}
\hline
\hline
$N_f$ & \# of $c, c^{\dagger}$ terms (Eq.~\eqref{eq:terms_GN_thirring}) & \# of Pauli strings (Eq.~\eqref{eq:Pauli_terms_GN_thirring}) & CX & Clifford+T \\ 
\hline
2  & 432 & 213 & 704 & 1960   \\
\hline 
4  & 1504  & 745 & 2124 & 4811  \\
\hline 
6  & 3216 & 1597 & 4474 & 8398   \\
\hline 
8  & 5568  & 2769 & 8022 & 13228  \\
\hline
\end{tabular}
\caption{Gate counts for varying number of flavors $N_f$ for fixed $L=10$ for GN model (top) and Thirring model (bottom) for one first-order Trotter step.}
\label{tab:cx_counts2}
\end{table}

\end{document}